\documentstyle{article}
\begin{document}

\title{\large \bf GENERALIZED STRONG CURVATURE SINGULARITIES \\
AND COSMIC CENSORSHIP}

\author{{\normalsize WIES{\L}AW RUDNICKI}\thanks{E-mail:
rudnicki@atena.univ.rzeszow.pl} \\
{\small \it Institute of Physics, University of Rzesz\'ow,
 Rejtana 16A, 35-959 Rzesz\'ow, Poland} \\
{\normalsize ROBERT J. BUDZY\'NSKI}\thanks{E-mail:
Robert.Budzynski@fuw.edu.pl} \\
{\small \it Department of Physics, Warsaw University,
 Ho\.za 69, 00-681 Warsaw, Poland} \\
{\normalsize WITOLD KONDRACKI}\thanks{E-mail:
witekkon@panim.impan.gov.pl} \\
{\small \it Institute of Mathematics, Polish Academy of Sciences,
 \'Sniadeckich 8, 00-950 Warsaw, Poland}}

\date{}
\maketitle
\begin{abstract}
A new definition of a strong curvature singularity is proposed. This
definition is motivated by the definitions given by Tipler and Kr\'olak,
but is significantly different and more general. All causal geodesics
terminating at these new singularities, which we call {\it generalized
strong curvature singularities,} are classified into three possible types;
the classification is based on certain relations between the causal structure
and the curvature strength of the singularities. A cosmic censorship theorem
is formulated and proved which shows that only one class of generalized
strong curvature singularities, corresponding to a single type of geodesics
according to our classification, can be naked. Implications of this
result for the cosmic censorship hypothesis are indicated.


\end{abstract}

\newpage

\section{Introduction}

The cosmic censorship hypothesis (CCH) of Penrose \cite{p1,p2} says that in
generic situations, all spacetime singularities arising from regular
initial data are always hidden behind event horizons and hence invisible
to outside observers (no naked singularities). This hypothesis plays a
fundamental role in the theory of black holes and has been recognized as
one of the most important open problems in classical general relativity.
There exist many exact solutions of Einstein's equations which admit
naked singularities. However, Penrose \cite{p2} has stressed that the exact
solutions with special symmetries have a rather limited value for
verification of the CCH and what is required here is an understanding of
the generic case.

One possible approach to this problem is to propose a class of generic
singularities and then attempt to formulate and prove a censorship theorem
which would constrain or prohibit the occurrence of naked singularities
of the proposed class. Tipler {\it et al.} \cite{t1} and Kr\'olak \cite{k1}
have argued that all singularities arising in generic situations should be of
the strong curvature type. These singularities have the property that all
objects approaching them are crushed to zero volume. The idea of a strong
curvature singularity was introduced by Ellis and Schmidt \cite{e1} and
defined in precise geometrical terms by Tipler \cite{t2} (a slightly different
definition was given by Kr\'olak \cite{k1}, see below). Kr\'olak \cite{k1,k2}
formulated and proved some censorship theorems that ruled out a class of
naked singularities of the strong curvature type. Unfortunately,
these results rely heavily on a further assumption (the so-called
simplicity condition) which need not hold for generic spacetimes.

It should be stressed that there is no hope for finding a proof of the
CCH using only the assumption that all singularities occurring in a given
spacetime are of the strong curvature type. This follows just from the fact
that naked singularities of this type {\it do} occur in certain exact
solutions of Einstein's equations. For instance, they occur in the
Tolman-Bondi solution representing spherically symmetric inhomogeneous
collapse of dust (see, e.g., Ref. \cite{d1}; see also Ref. \cite{j1} and
references therein).
Such naked singularities also occur in more general models of dust collapse
--- namely, in the Szekeres spacetimes which do not have any Killing
vectors \cite{k3,j2}. Unnikrishnan \cite{u1} has argued that the existence
of naked strong curvature
singularities in the Tolman-Bondi solution can be ruled out by imposing
certain reasonable constraints on the initial distribution of the energy
density of dust. This argument, however, depends crucially on the spherical
symmetry of the solution, and so cannot be applied to naked singularities
occurring in more general cases --- e.g. in the Szekeres spacetimes. It is
worth recalling here that Bonnor \cite{b1} remarked, in another context, that
``the Szekeres solution has a good deal of symmetry, even though it has in
general no Killing vectors.'' It is thus possible that the existence of naked
singularities of the strong curvature type will always be accompanied by
spacetime symmetries and/or instabilities of some sort, and so one can still
hope to prove a formulation of the CCH involving --- besides the assumption
that all singularities are of strong curvature --- a suitable criterion of
genericity or stability.

To help identify such a criterion, it may be useful to establish and
study various relations between strong curvature singularities and the
causal structure in their neighborhood --- such relations get to the
heart of cosmic censorship. For this purpose, we need a new definition
of a strong curvature singularity --- one that not only describes the
curvature strength of the singularity, but additionally enables one to
relate the strength with properties of the causal structure in a
neighborhood of the singularity. In Section II of this paper, we shall
propose such a definition. Our definition is motivated by the definition
of a strong curvature singularity given by Tipler and its modifications by
Kr\'olak, but is significantly different and more general. The difference is
in that our definition involves a certain focusing condition on solutions of
the Raychaudhuri equation not only along a {\it single} causal geodesic, as
in Tipler's and Kr\'olak's case, but rather along {\it all} causal geodesics
in some small neighborhood about a given geodesic that reaches the
singularity. All causal geodesics terminating at the singularities
described by our definition, which we will refer to as {\it generalized
strong curvature singularities}, will next be classified into three possible
types; the classification is based on certain relations between the causal
structure and the curvature strength of the singularities. Further on, in
Sections III and IV, we shall formulate and prove a cosmic censorship theorem
which shows that only one class of generalized strong curvature singularities,
corresponding to a single type of geodesics according to our classification,
can be naked.

This paper was inspired by ideas given in Ref. \cite{r1}. Some of the results
presented here are refinements of those announced without proofs in Ref.
\cite{k4}.
The notation and fundamental definitions are as those in the monograph of
Hawking and Ellis \cite{h1}.

\section{Generalized Strong Curvature Singularities}

Before we give our definition, we need to recall one standard result on
the behavior of geodesic congruences. Let $\lambda(t)$ be an affinely
parametrized null (resp. timelike) geodesic. A congruence of null (timelike)
geodesics infinitesimally neighboring $\lambda(t)$ and originating from a
point on $\lambda(t)$ is characterized by two parameters: the expansion
$\theta$ and the shear $\sigma$. The rate of change of the expansion
$\theta$ along $\lambda(t)$ is given by the Raychaudhuri equation:
\begin{equation}
\frac{d\theta }{dt}=-R_{ab}K^{a}K^{b}-2\sigma ^{2}-\frac{1}{n}\theta ^{2},
\end{equation}
where $R_{ab}$ is the Ricci tensor, $K^{a}$ is the tangent vector to
$\lambda(t)$, $n=2$ if $\lambda(t)$ is a null geodesic and $n=3$ in the
case when it is timelike (see pp. 78--88 of Ref. \cite{h1}).

Let us also recall the definition of a strong curvature singularity
introduced by Kr\'olak \cite{k1}.

\vspace{0.3in}
\noindent {\bf Definition 1:} {\em Let $\lambda$ be a future-endless,
future-incomplete null (timelike) geodesic. $\lambda$ is said to terminate
in the future at a strong curvature singularity if, for each point $p\in
\lambda$, the expansion $\theta$ of every future-directed congruence of
null (timelike) geodesics emanating from $p$ and containing $\lambda$
becomes negative somewhere on $\lambda$.}

\vspace{0.3in}
This definition is equivalent to condition K given in Ref. \cite{c1}.
Kr\'olak's definition generalizes Tipler's definition of a strong
curvature singularity because it implies weaker restrictions on the
divergence of the curvature near the singularity \cite{c1}. 

We can now introduce our definition of a generalized strong curvature
singularity.

\vspace{0.3in} \noindent
{\bf Definition 2:} {\em Let $\lambda$ be a future-endless,
future-incomplete null (timelike) geodesic. We say that $\lambda$
terminates in the future at a generalized strong curvature singularity if,
for each sequence of endless null (timelike) geodesics, $\{\lambda_{n}\}$,
for which $\lambda$ is a limit curve, at least one of
the following conditions holds:}
\begin{description}
\item[(i)] {\em for each point $p\in \lambda$ and for each neighborhood $N$
of $p$ there exist a geodesic $\widetilde{\lambda}\in \{\lambda_{n}\}$ and a
point $\widetilde{q}\in \widetilde{\lambda}\cap N$ such that the
expansion $\theta$ of a future-directed congruence of null (timelike)
geodesics emanating from $\widetilde{q}$ and containing
$\widetilde{\lambda}$ becomes negative somewhere on $\widetilde{\lambda}$;}
\item[(ii)] {\em $I^{-}(\lambda)=I^{-}(\lambda_{n})$ for almost all
geodesics belonging to
$\{\lambda_{n}\}$}.
\end{description}
\noindent
{\em Analogously, $\lambda$ is said to terminate in the past at a generalized
strong curvature singularity if the time-reverse versions of the above
conditions hold for $\lambda$.}

\vspace{0.3in}
\noindent {\bf Remark:} By the term {\it limit curve} we mean in the above
a curve that satisfies the definition given on p. 185 of Ref. \cite{h1}.

It is easy to notice that every causal geodesic that terminates at a
strong curvature singularity as defined by Kr\'olak also terminates at
a generalized strong curvature singularity as defined above. To see this,
consider, e.g., a null geodesic $\lambda$ that terminates in the future
at Kr\'olak's singularity. Next, take any sequence of endless null
geodesics, $\{\lambda_{n}\}$, for which $\lambda$ is a limit curve. Let
$p$ be an arbitrary point on $\lambda$, and let $\{p_{n}\}$ be an arbitrary
sequence of points converging to $p$ such that, for each $n$, $p_{n}\in
\lambda_{n}$. In accordance with Kr\'olak's definition, the expansion
$\theta$ of every future-directed congruence of null geodesics outgoing
from $p$ and containing $\lambda$ must become negative. It follows by
continuity that there must exist certain points $p_{k}\in \{p_{n}\}$
sufficiently close to $p$\, such that, for each $k$, the expansion
$\theta$ of a future-directed congruence of null geodesics outgoing from
$p_{k}$ and containing $\lambda_{k}\in \{\lambda_{n}\}$ shall become
negative as well. This clearly means that $\{\lambda_{n}\}$ shall obey
condition (i) of Definition 2.

Let us now turn to condition (ii) of Definition 2. This condition has no
direct equivalent in either Tipler's or Kr\'olak's definition; however,
it appeals to the original idea behind the concept of a strong curvature
singularity \cite{e1}. To see this, observe first that the requirement
$I^{-}(\lambda)=I^{-}(\lambda_{n})$ found in condition (ii) means that
both $\lambda$ and $\lambda_{n}$ must reach exactly the same point in the
{\em c-boundary} of the spacetime (p. 218 of Ref. \cite{h1}). The geodesics
$\lambda_{n}$ represent trajectories of point particles in motion towards
the singularity within some small neighborhood about $\lambda$. Following
the original idea of a strong curvature singularity, all small physical
objects should be crushed to zero volume as they approach the singularity.
Condition (ii) corresponds to the particular case where they are crushed
to a single point in the $c$-boundary.

It is worth mentioning here that the $c$-boundary of cosmological models
has been investigated by Tipler \cite{t3} in the context of limitations
imposed on computation by general relativity. He has shown that a true
universal Turing machine can be constructed only in a closed universe
whose final singularity is a single point in the $c$-boundary topology.
Thus all causal geodesics reaching this singularity will satisfy condition
(ii) of Definition 2 and hence terminate in our generalized strong
curvature singularity; moreover, they will be of type A according to the
classification given below.

The following definition provides a complete classification of causal
geodesics terminating at the singularities introduced in Definition 2.

\vspace{0.3in} \noindent
{\bf Definition 3:} {\em Let $\lambda$ be a future-endless null
(timelike) geodesic terminating in the future at a generalized strong
curvature singularity, and let $\{\lambda_{n}\}$ be a sequence of
endless null (timelike) geodesics for which $\lambda$
is a limit curve. $\lambda$ is said to be of type:}
\begin{itemize}
\item {\em $A$, if condition (2) of Definition 2 holds for each
$\{\lambda_{n}\}$;}
\item {\em $B$, if, for each $\{\lambda_{n}\}$ which does not satisfy
condition (2) of Definition 2, there exist a geodesic
$\widetilde{\lambda}\in \{\lambda_{n}\}$ and a point $\widetilde{q}\in
\widetilde{\lambda}-I^{-}(\lambda)$ such that the expansion $\theta$ of a
future-directed congruence of null (timelike) geodesics emanating from
$\widetilde{q}$ and containing $\widetilde{\lambda}$ becomes negative
somewhere on $\widetilde{\lambda}$;}
\item {\em $C$, if $\lambda$ is neither of type $A$ nor $B$.}
\end{itemize}

\vspace{0.3in} \noindent
{\bf Remark:} If the $\lambda$ mentioned above is a null geodesic that admits
a segment contained in the boundary, $\dot{I}^{-}(\lambda)$, of its
chronological past, one can always find a sequence of endless null geodesics,
$\{\lambda_{n}\}$, for which $\lambda$ is a limit curve, such that none of
the $\lambda_{n}$ will be contained in the closure of $I^{-}(\lambda)$.
Clearly, such a sequence cannot satisfy condition (ii) of Definition 2, and
so $\lambda$ cannot be of type A. In general, however, there may exist a
future-endless null geodesic $\lambda$ that never intersects the boundary of
its chronological past. For example, if $\lambda$ terminates in the future
at a curvature singularity and the curvature diverges fast enough
along $\lambda$, every future-endless segment of $\lambda$ may admit a pair
of conjugates points, which implies that no segment of $\lambda$ can be
contained in the achronal boundary $\dot{I}^{-}(\lambda)$. In this case
$\lambda$ may be of type A.

\section{Censorship Theorem}

For the sake of simplicity, we shall restrict our considerations to {\it
weakly asymptotically simple and empty} (WASE) spacetimes (p. 225 of Ref.
\cite{h1}). Such a spacetime $(M,g)$ can be conformally imbedded in a larger
spacetime $(\widetilde{M},\widetilde{g})$ as a manifold with boundary
$\overline{M}=M\cup \partial M$, where the boundary $\partial M$ consists
of two null surfaces ${\cal J}^{+}$ and ${\cal J}^{-}$ that represent
future and past null infinity, respectively. Moreover, there exists an open
neighborhood $U$ of $\partial M$ in $\widetilde{M}$ such that $U\cap M$
coincides with part of an asymptotically simple and empty spacetime
$(M',g')$, which means that all possible singularities of $(M,g)$ can only
occur in the region $M-U$. Since the CCH is concerned with singularities
that develop from an initially non-singular state, we shall deal here only
with such WASE spacetimes $(M,g)$ in which the region $M-U$ does not
extend arbitrarily far into the past and to a spatial infinity.\footnote{Due
to its generality, the definition of a WASE spacetime given in Ref. \cite{h1}
has the unwanted feature that the region $M-U$ might extend to a spatial
infinity. For a deeper discussion of this problem, see Ref. \cite{c1}.}
To make this precise, we shall assume that $(M,g)$ admits a partial Cauchy
surface $S$ for which the following two conditions hold:
\begin{description}
\item[(i)] {\em $S$ has an asymptotically simple past} (p. 316 of Ref.
\cite{h1});
\item[(ii)] {\em every null geodesic $\mu$ generating ${\cal J}^{+}$ admits
a past-endless segment $\mu_{0}$ such that $\mu_{0}\subset
I^{+}(S,\overline{M})$ and $I^{-}(\mu_{0},\overline{M})\cap
I^{+}(S)\subset D^{+}(S)\cap U$.}
\end{description}
\noindent
Such a surface $S$ will be called a {\it regular} partial Cauchy surface.
(This definition is a slight modification of that previously used in Ref.
\cite{r2}.)

Let us now recall that a WASE spacetime $(M,g)$ is said to be {\it
future asymptotically predictable} from a partial Cauchy surface $S$ if
the future null infinity ${\cal J}^{+}$ is contained in the closure of the
future Cauchy development $D^{+}(S,\overline{M})$ (p. 310 of Ref. \cite{h1}).
Future asymptotic predictability is a mathematically precise statement of
cosmic censorship for $(M,g)$, since it ensures that there will be no
singularities to the future of $S$ which are naked, i.e. which are visible
from ${\cal J}^{+}$.

We are now in a position to state our cosmic censorship theorem. By this
theorem, only singularities corresponding to null geodesics of type C can be
naked.

\vspace{0.3in} \noindent
{\bf Theorem 1:} {\em Let $(M,g)$ be a WASE spacetime admitting a
regular partial Cauchy surface $S$. Suppose furthermore that the
following conditions hold:}
\begin{description}
\item[(i)] {\em the null convergence condition, i.e. $R_{ab}K^{a}K^{b}\geq 0$
for every null vector $K^{a}$ of $(M,g)$;}
\item[(ii)] {\em the generic condition, i.e. every endless null geodesic
of $(M,g)$ admits a point at which $K_{[a}R_{b]cd[e}K_{f]}K^{c}K^{d}\neq
0$, where $K^{a}$ is the tangent vector to the geodesic;}
\item[(iii)] {\em $(M,g)$ admits no naked points-at-infinity, i.e. for each
point $p\in {\cal J}^{+}$, every future-endless, future-complete null
geodesic of $(M,g)$ contained in $\bar{I}^{-}(p,\overline{M})\cap D(S)$
has a future endpoint on ${\cal J}^{+}$ in $\overline{M}$;}
\item[(iv)] {\em every incomplete null geodesic of $(M,g)$ terminates at a
generalized strong curvature singularity.}
\end{description}
\noindent
{\em If $(M,g)$ is not future asymptotically predictable from $S$, then
there must exist a null geodesic $\lambda\subset$}\, int{\em $D(S)$
which terminates in the future at a generalized strong curvature
singularity and is of type C.}

\vspace{0.3in} \noindent
{\bf Remarks:} Conditions (i) and (ii) of Theorem 1 are reasonable
requirements for {\em any} physically realistic model of a classical
spacetime; they have been discussed extensively in the literature on
the singularity theorems (see, e.g., Refs. \cite{t1,h1}).

Condition (iii) is a slightly weakened version of that previously used by
Newman and Joshi \cite{n1} in their censorship theorem. This condition
ensures that any possible breakdown of future asymptotic predictability in
$(M,g)$ must be associated with the occurrence of a naked singularity and not
of a naked {\em ``point-at-infinity''}. Known examples of WASE spacetimes
with naked singularities contain no naked points-at-infinity in the
sense of condition (iii). It should, however, be stressed that this
condition restricts to some extent the generality of our result, since it
is conceivable that some spacetimes might contain both naked singularities
and naked points-at-infinity.

Note also that no causality conditions, other than those implied by the
existence of $S$, are imposed on $(M,g)$ in Theorem 1. Thus this
theorem may be applied to naked singularities associated with causality
violation. The possible existence of such naked singularities results
from Tipler's singularity theorem \cite{t4}.

\section{Proof of the Theorem}

Now we shall prove our theorem; the following two lemmas serve this
purpose.

\vspace{0.3in} \noindent
{\bf Lemma 1:} {\em Under the assumptions of Theorem 1, suppose that
$(M,g)$ is not future asymptotically predictable from $S$. Then there
must exist a past-incomplete null geodesic $\eta\subset
H^{+}(S,\overline{M})$ which has a future endpoint $p\in {\cal J}^{+}$.
Moreover, the following conditions hold:}
\begin{description}
\item[(a)] {\em there exists a point $r\in \eta\cap M$ such that
the closure of $I^{-}(r)\cap S$ is compact;}
\item[(b)] {\em for each point $c_{i}\in I^{-}(p,\overline{M})\cap D^{+}(S)$,
there exists a null geodesic $\lambda_{i}\subset
\dot{I}^{+}(c_{i},\overline{M})$ which extends from $c_{i}$ to some
point on ${\cal J}^{+}$.}
\end{description}

\vspace{0.3in} \noindent
{\bf Proof:} Let us first observe that from condition (ii) of the
definition of $S$ it follows that every generator of
${\cal J}^{+}$ must intersect the closure of $D^{+}(S,\overline{M})$.
In addition, as $S$ has an asymptotically simple past by condition (i),
${\cal J}^{-}$ must be contained in the closure of $D^{-}(S,\overline{M})$.
This implies that $(M,g)$ is {\em partially} asymptotically predictable from
$S$ as defined by Tipler \cite{t4}. Thus, as $(M,g)$ is not future
asymptotically predictable from $S$, by Proposition 2 of Ref. \cite{t4} there
must exist a past-endless null geodesic generator $\eta$ of $H^{+}(S)$ with
future endpoint $p\in {\cal J}^{+}$. Following steps as in the proof of
Theorem 1 of Ref. \cite{t4}, we can now easily show, by making use of
conditions (i) and (ii) of our theorem, that $\eta$ must be incomplete in
the past.

We shall now show that condition (a) holds. Let us choose an arbitrary
point $r\in \eta\cap M$; we shall demonstrate that the set $R\equiv
\overline{I^{-}(r)\cap S}$ is compact. Since $S$ has an asymptotically
simple past, the proof of this fact will be quite similar to the proof
of the Theorem of Ref. \cite{k5}.

Assume, to the contrary, that $R$ is not compact. Then, by Lemma 2 of Ref.
\cite{k5}, there must exist a sequence of null geodesics $\eta_{i}$ with
future endpoints $r_{i}$ converging to $r$, such that each $\eta_{i}$ will
be a generator of the achronal boundary $\dot{I}^{-}(r_{i})$ and will have
a past endpoint on ${\cal J}^{-}$; moreover, the geodesic $\eta$ must be a
limit curve of the sequence $\{\eta_{i}\}$. As $\eta$ is past-incomplete,
by condition (iv) of Theorem 1, it must terminate in the past at a
generalized strong curvature singularity. Thus $\{\eta_{i}\}$ must fulfill
at least one of the time-reverse versions of conditions (i) and (ii) of
Definition 2. Since each of the $\eta_{i}$ has a past endpoint on
${\cal J}^{-}$, for each $\eta_{i}$ one has $I^{+}(\eta_{i})\neq
I^{+}(\eta)$, as $\eta \subset I^{+}(S)$ while ${\cal J}^{-}\subset
I^{-}(S,\overline{M})$. It is thus clear that the time-reverse version of
condition (ii) cannot be fulfilled, and so the time-reverse version of
condition (i) must be. There must thus exist a geodesic $\widetilde{\eta}\in
\{\eta_{i}\}$ and a point $\widetilde{q}\in \widetilde{\eta}$ such that the
expansion $\theta$ of a past-directed null geodesic congruence emanating
from $\widetilde{q}$ and containing $\widetilde{\eta}$ must become negative
somewhere on $\widetilde{\eta}$. In this case, by Proposition 4.4.4 of Ref.
\cite{h1} and condition (i) of our Theorem 1, there would exist a point
conjugate to $\widetilde{q}$ along $\widetilde{\eta}$, since
$\widetilde{\eta}$ is past-complete as it has a past endpoint on
${\cal J}^{-}$. However, by Proposition 4.5.12 of Ref. \cite{h1}, the
existence of a pair of conjugate points on $\widetilde{\eta}$ would
contradict the achronality of $\widetilde{\eta}$.
In view of this contradiction we must conclude that $R$ is compact.

We shall now show that condition (b) holds. Let $c_{i}$ be an arbitrary
point in $I^{-}(p,\overline{M})\cap D^{+}(S)$, and let $\mu$ be a
generator of ${\cal J}^{+}$ that passes through $p$. We shall first show
that $\mu$ must leave $I^{+}(c_{i},\overline{M})$. To see this, suppose
that $\mu$ were contained in $I^{+}(c_{i},\overline{M})$. Then the past
set $X\equiv \bigcap_{a\in\mu_{0}}[I^{-}(a,\overline{M})\cap M]$, where
$\mu_{0}$ denotes the past-endless segment of $\mu$ mentioned in condition
(ii) of the definition of $S$, would be non-empty, as it would contain
$c_{i}$. From the definition of $X$ it follows that any null geodesic
generator of the boundary of $X$ cannot have its future endpoint on
${\cal J}^{+}$; otherwise, such a point would have to coincide with a past
endpoint of $\mu_{0}$, which is impossible as $\mu_{0}$ is past-endless.
But from the definition of the surface $S$ it follows that $X$ would have to
be contained in an open neighborhood $U$ of ${\cal J}^{+}\cup {\cal J}^{-}$,
such that $U\cap M$ coincides with part of an asymptotically simple and
empty spacetime, which implies that all null geodesics generating the
boundary of $X$ would have to have their future endpoints on ${\cal J}^{+}$.
This contradiction shows that $\mu$ must leave $I^{+}(c_{i},\overline{M})$,
so there must exist a point $b\in \mu\cap\dot{I}^{+}(c_{i},\overline{M})$.
Since $b\in \mu \cap J^{-}(p,\overline{M})$, and $p\in H^{+}(S,\overline{M})$,
it follows that $b$ must belong to the closure of $D^{+}(S,\overline{M})$.
Therefore all the past-directed null geodesics outgoing from $b$, with
the exception of $\mu$, must enter the interior of $D^{+}(S)$ immediately
after leaving $b$; otherwise they would be generators of ${\cal J}^{+}$,
which is impossible. Thus, as int$D^{+}(S)$ is a causally simple set, there
must exist a null geodesic $\lambda_{i}\subset
\dot{I}^{+}(c_{i},\overline{M})$ which extends from $c_{i}$ to $b\in
\cal{J}^{+}$, as required in condition (b). \hfill $\Box$ \\

\vspace{0.3in} \noindent
{\bf Lemma 2:} {\em Let $r$ be a point on $H^{+}(S)$, where $S$ is a
partial Cauchy surface. If the set $\dot{I}^{-}(r)\cap J^{+}(S)$ is compact,
then every null geodesic generator of $H^{+}(S)$ through $r$ is geodesically
complete in the past direction.}

\vspace{0.3in} \noindent
{\bf Proof:} The proof of this lemma is identical to the proof of Lemma 8.5.5
of Ref. \cite{h1}. In the course of the proof we only need to consider the
set $\dot{I}^{-}(r)\cap J^{+}(S)$ instead of the Cauchy horizon $H^{+}(S)$.
\hfill $\Box$

\vspace{0.3in} \noindent
{\bf Proof of Theorem 1:} Assume that $(M,g)$ is not future asymptotically
predictable from $S$. Then, by Lemma 1, there exists a past-incomplete
null geodesic $\eta\subset H^{+}(S)$ which has a future endpoint $p\in
{\cal J}$; moreover, there exists a point $r\in \eta\cap M$ such that
$R\equiv \overline{I^{-}(r)\cap S}$ is compact. As $\eta$ is past-incomplete,
by Lemma 2 the set $Q\equiv \dot{I}^{-}(r)\cap J^{+}(S)$ cannot be compact.
Let us put a timelike vector field on $M$ (such a field will always exist
since $M$ admits a Lorentz metric $g$). If every integral curve of this
field that intersects the set $R$ also intersects the set $Q$, we would have
a continuous one-to-one mapping of $R$ onto $Q$, and hence $Q$ would have to
be compact. Therefore there must exist a future-endless timelike curve
$\alpha\subset I^{-}(r)$.

Let ${\cal P}$ be the family of all sets of the form $I^{-}(\alpha)$,
where $\alpha$ is a future-endless timelike curve contained in
$I^{-}(r)$; and let $\widehat{P}$ be a maximal chain determined in
${\cal P}$ by the relation of inclusion. Denote now by $P_{0}$ the set
$\bigcap\{P: P\in \widehat{P}\}$. This set is non-empty. To see this, note
first that no member of $\widehat{P}$ can be contained in $I^{-}(S)$;
otherwise, there would exist a future-endless timelike curve
$\alpha\subset I^{-}(S)$, which is impossible as $I^{-}(S)=D^{-}(S)$.
Note now that, for each $P\in \widehat{P}$, the set $\overline{P\cap S}$
must be compact, as it is a closed subset of the compact set $R$. This
clearly implies, by the definition of $P_{0}$, that the set $A\equiv
\overline{P_{0}\cap S}$ is non-empty and compact. In addition, from the
definition of $P_{0}$ it follows that there must exist a future-endless
timelike curve $\alpha_{0}$ such that $P_{0}=I^{-}(\alpha_{0})$. Note also
that $P_{0}$ must be a minimal element of $\widehat{P}$, i.e. we will have
$P_{0}=I^{-}(\beta)$ for every future-endless timelike curve
$\beta\subset P_{0}$.

Let us now fix some future-endless timelike curve $\alpha_{0}$ such
that $P_{0}=I^{-}(\alpha_{0})$. Let $\{c_{i}\}$ be a sequence of points
on $\alpha_{0}\cap D^{+}(S)$ such that, for each $i$, $c_{i+1}\in
I^{+}(c_{i})$; assume also that $\{c_{i}\}$ has no accumulation point on
$\alpha_{0}$ (such a sequence can always be found as $\alpha_{0}$ has no
future endpoint). By condition (b) of Lemma 1, there must exist, for each
$i$, a null geodesic $\lambda_{i}$ running from $c_{i}$ to some point on
${\cal J}^{+}$. Moreover, each $\lambda_{i}$ shall be a generator of the
achronal boundary $\dot{I}^{+}(c_{i},\overline{M})$ and shall be
future-complete as it reaches ${\cal J}^{+}$.

Let us now extend each of the $\lambda_{i}$ maximally into the past.
Since each $\lambda_{i}$ passes through $c_{i}\in \alpha_{0}\cap
D^{+}(S)$, each of the extended $\lambda_{i}$ must intersect $S$ at some
point $a_{i}\in A$. As the set $A$ is compact, the sequence $\{a_{i}\}$
must have an accumulation point $a\in A$. Therefore, by Lemma 6.2.1 of
Ref. \cite{h1}, through $a$ there is a non-spacelike curve $\lambda$ which is
future-inextendible in $M$ and which is a limit curve of the sequence
$\{\lambda_{i}\}$. Since all the $\lambda_{i}$ are null geodesics,
$\lambda$ must be a null geodesic as well. Note also that $\lambda$ must be
contained in $\overline{P_{0}}$; otherwise $\lambda$ would have to intersect
the curve $\alpha_{0}$ at some point that would be an accumulation point of
the sequence $\{c_{i}\}$, which is impossible as $\{c_{i}\}$ has no
accumulation point on $\alpha_{0}$. Moreover, as $P_{0}$ is a minimal element
of $\widehat{P}$, we must have $P_{0}=I^{-}(\lambda)$. Since $\lambda$
intersects the surface $S$, it cannot be a generator of $H^{+}(S)$. Thus, as
$\lambda\subset\overline{P_{0}}\subset\overline{D(S)}$, we will have
$\lambda\subset$ int$D(S)$. Since $\lambda\subset
\overline{P_{0}}\subset \overline{I^{-}(r)}$, and $r\in M$, $\lambda$
cannot have a future endpoint on $\cal{J}^{+}$. In addition, as
$r\in J^{-}(p,\overline{M})$, $\lambda$ must be contained in the closure
of $I^{-}(p,\overline{M})$. Therefore, as $p\in {\cal J}^{+}$, by
condition (iii) of Theorem 1, $\lambda$ must be incomplete in the future.

By condition (iv) of Theorem 1, $\lambda$ must terminate in the future at a
generalized strong curvature singularity. Thus the sequence $\{\lambda_{i}\}$
must satisfy at least one of conditions (i) and (ii) of Definition 2. Since
$\lambda\subset\overline{P_{0}}$, and each of the $\lambda_{i}$ must leave
$\overline{P_{0}}$ as it has a future endpoint on ${\cal J}^{+}$, we will
have $I^{-}(\lambda)\neq I^{-}(\lambda_{i})$ for each $i$. This means that
condition (ii) of Definition 2 cannot hold for the $\{\lambda_{i}\}$, hence
$\lambda$ cannot be of type A. Suppose that $\lambda$ were of type B. Then,
according to Definition 3, there would exist a geodesic $\widetilde{\lambda}
\in \{\lambda_{i}\}$ and a point $\widetilde{q}\in
\widetilde{\lambda}-I^{-}(\lambda)$ such that the expansion $\theta$ of a
future-directed congruence of null geodesics outgoing from $\widetilde{q}$
and containing $\widetilde{\lambda}$ would be negative somewhere on
$\widetilde{\lambda}$. As $\widetilde{\lambda}$ is future-complete, by
condition (i) of our theorem and Proposition 4.4.4 of Ref. \cite{h1}, there
would thus exist some point $x\in J^{+}(\widetilde{q})$ conjugate to
$\widetilde{q}$ along $\widetilde{\lambda}$. Consequently, by Proposition
4.5.12 of Ref. \cite{h1}, there would also exist a timelike curve from
$\widetilde{q}$ to some point $y\in \widetilde{\lambda}\cap J^{+}(x)$. But
this is impossible because from the above construction it follows that the
points $\widetilde{q}$ and $y$ would have to lie on the achronal segment of
$\widetilde{\lambda}$ contained in the boundary
$\dot{I}^{+}(\alpha_{0}\cap\widetilde{\lambda})$. This contradiction shows
that the geodesic $\lambda$ cannot be of type B. Thus, as $\lambda$
is neither of type A nor B, it must be of type C, which completes the proof.
\hfill $\Box$

\section{Concluding Remarks}
In Theorem 1 we have demonstrated that the only possible naked singularities
of strong curvature are those corresponding to null geodesics of type C, 
according to the classification proposed in this paper. As a consequence, one 
only needs to consider the stability of type C singularities as relevant to
the cosmic censorship problem, which is a significant restriction. One
argument showing that these singularities have a tendency to be unstable
under generic perturbations was given in Ref. \cite{k4}. We may conclude that
a further study of the properties of type C singularities might bring us
closer to formulating the genericity and stability criterium which is needed,
as mentioned in the Introduction, to arrive at a satisfactory statement of
the CCH.

As we mentioned above, naked strong curvature singularities are present in
certain exact solutions of the Einstein equations --- solutions that represent
gravitational collapse. In view of our result, it would therefore be of
interest to verify whether in those solutions, the naked singularity is also
associated in every case with the existence of type C geodesics. It should
be stressed that this does not follow directly from Theorem 1, since it is
not clear whether the solutions in question necessarily satisfy all the
assumptions of that theorem.

\section*{Acknowledgements}

We wish to thank Andrzej Kr\'olak for many valuable discussions. This
work was supported in part by the Polish Committee for Scientific
Research (KBN) under Grant No. 2 P03B 130 16.

\end{document}